\documentclass[epj,nopacs]{svjour}

\usepackage{graphicx}% Include figure files

\newcommand{\bq}{\begin{equation}}
\newcommand{\eq}{\end{equation}} 
\newcommand{\ba}{\begin{eqnarray}}
\newcommand{\ea}{\end{eqnarray}}

\begin{document}
\title{Cosmological Constraints on Unparticle Dark Matter} 
\author{Yan Gong \and Xuelei Chen}
%\offprints{Xuelei Chen}
\institute{National Astronomical Observatories, Chinese Academy of Sciences,\\
 20A Datun Rd, Chaoyang District, Beijing 100012, China}
\date{Received: March 20th, 2008}

\abstract{
In unparticle dark matter (unmatter) models the equation of state 
of the unmatter is given by $p=\rho/(2d_U+1)$, where $d_U$ is 
the scaling factor. Unmatter with such equations of state would have a
significant impact on the expansion history of the universe. Using
type Ia supernovae (SNIa), the baryon acoustic oscillation (BAO) 
measurements and the shift parameter of the cosmic microwave 
background (CMB) to place
constraints on such unmatter models we find that if only the SNIa data is used
the constraints are weak. However, with the BAO and CMB shift parameter data added
strong constraints can be obtained. For the $\Lambda$UDM model, in which
unmatter is the sole dark matter, we find that $d_U > 60$ at 95\%
C.L. For comparison, in most unparticle physics models it is assumed $d_U<2$. 
For the $\Lambda$CUDM model, in which unmatter co-exists with cold
dark matter, we found that the unmatter can at most make up a few 
percent of the total cosmic density if $d_U<10$, thus it can not be the 
major component of dark matter. 
}

\maketitle

\section{Introduction}

It has recently been proposed that a hidden scale-invariant sector 
of matter may exist \cite{Georgi07}, and is named ``unparticle'' for 
its unusual behavior. In this scenario, there is a scale invariant sector 
with a non-trivial infrared fixed point, called the Banks-Zaks (BZ) 
field \cite{Banks82}. The BZ field interacts with the standard model
(SM) fields via exchange of particles of mass $M_U$:
$${\mathcal L_{BZ}}= \frac{O_{BZ}O_{SM}}{M_U^k},$$
where $O_{BZ}$ is the BZ operators with mass dimension $d_{BZ}$ 
and $O_{SM}$ is the SM operators with mass dimension
$d_{SM}$. Dimensional transmutation produce a scale $\Lambda_U$, below
which 
$${\mathcal L}_{BZ} \to {\mathcal L}_U = 
C_U \frac{\Lambda_U^{d_{BZ}-d_U}}{M_U^k} O_{SM}O_{U},$$
where $C_U$ is the coefficient function and $O_{U}$ is 
the unparticle operators with mass dimension $d_{U}$.    
Phenomenological constraints on $M_U,\Lambda_U,C_U,d_U$ have been 
derived from a number of particle physics
\cite{Georgi07b,Cheung07,Luo07,ChenGeng,Ding07,ACG07,Liao07,Li07,0705,Zhou07,Lu07,ChenHe,Huang,KO,Balantekin07}
and astrophysics
\cite{Davoudiasl07,Hannestad07,DHJ07,AG07,McDonald07,Das07,Freitas07}
analysis.

Given that the unparticles interacts weakly with standard model
particles, it is natural to consider unparticle matter, or 
{\it unmatter}, as a candidate of dark matter, especially if the unparticle
could be stabilized by a discrete symmetry
\cite{Kikuchi07}. Thanks to the unusual kinematics of unparticles,
the behavior of unmatter is distinctly different 
from the usual cold dark matter.

The unparticles do not have a fixed mass, and the density of states of 
unparticle is given by 
\bq
\label{eq:dos}
\frac{d^4p}{(2\pi)^4} 2p^0 \theta(p^0) \theta(p^2) (p^2)^{d_U-2}.
\eq
Typically $1<d_U<2$ are considered \cite{Georgi07,Banks82,Georgi07b,Cheung07,Luo07,ChenGeng,Ding07,ACG07,Liao07,Li07,0705,Zhou07,Lu07,ChenHe,Huang,KO,Balantekin07,Davoudiasl07,Hannestad07,DHJ07,AG07,McDonald07,Das07,Freitas07}. 
For a thermal distribution of unparticles, the density and pressure is
given by \cite{Chen07}
\begin{eqnarray}
p_U&=&g_s T^4 \left(\frac{T}{\Lambda}_U\right)^{2d_U-1} 
\frac{{\mathcal{C}(d_U)}}{4\pi^2}\\
\rho_U &=& (2d_U+1)g_s T^4 \left(\frac{T}{\Lambda}_U\right)^{2d_U-1} 
\frac{{\mathcal{C}(d_U)}}{4\pi^2}
\end{eqnarray}
where ${\mathcal C}(d_U)=B(3/2,d_U) \Gamma(2d_U+2) \zeta(2d_U+2)$, and 
$B,\Gamma,\zeta$ are the Beta, Gamma and Zeta functions. The equation of
state for the unmatter is therefore
\bq
w_U = 1/(2d_U+1)
\eq
Thus, as the universe expands, the energy density of unmatter 
evolves as $\rho_U(z)=\rho_{U0}(1+z)^{3(1+w_U)}$.
If $d_U= 1$, this is the same as radiation, and for $d_U \to \infty$ 
its behavior would be similar to cold dark matter. In the intermediate
case, its evolution would differ from 
both radiation and cold dark matter. We can then use 
cosmological observations to constrain the value of $d_U$. 

We shall consider two models. In both cases we assume the universe
is flat with a cosmological constant. In the first case, denoted by
$\Lambda$UDM, the unmatter serves as the sole dark matter, and the 
set of cosmological parameters $\theta$ is  
$\{\Omega_{b0},~\Omega_{U0},~d_U,~h_0~\}.$ In the second case, 
denoted by $\Lambda$CUDM, we consider the more general case where 
the unmatter is not the only source of dark matter.
We then constrain the amount of unmatter if it does exist. 
The cosmological parameters set $\theta$ in this case is
$\{ \Omega_{m0},~\Omega_{U0},~d_U,~h_0\}$. We then use a Markov Monte Carlo 
Chain method to make the global fitting and constraints. For details of 
our MCMC code we refer the readers to Ref.~\cite{Gong07}.

We consider three observational constraints. The first one is luminosity
distance moduli to type Ia supernovae (SNIa). The second is the 
baryon acoustic oscillation (BAO) feature in large scale structure as 
measured by
the Sloan Digital Sky Survey (SDSS) and 
the Two Degree Field Galaxy Redshift Survey (2dFGRS).
The last one is the so called shift
parameter\cite{Wang06,Wang07}, which is essentially a measure of the
distance to the last scattering surface of the cosmic microwave
background (CMB), as measured by the WMAP three year observation
\cite{Spergel06}. 
All of these provide constraints on the global expansion 
history of the universe.

\section{Methods}

The cosmic expansion rate $H(z)$ is given by
\bq 
H^2(z) = H_0^2\ \Omega({\bf z;\theta}) ,
\eq
where for $\rm \Lambda UDM$:
\ba
\Omega({\bf z;\theta}) &=& 
\Omega_{b0} (1+z)^3 + \Omega_{\Lambda 0}+\Omega_{r0} (1+z)^4\nonumber\\ 
&&+ \Omega_{U0} (1+z)^{3(1+w_U)}
\ea
and for $\rm \Lambda CUDM$:
\ba
\Omega({\bf z;\theta}) &=& \Omega_{m_0} (1+z)^3 
+ \Omega_{\Lambda 0} +\Omega_{r0} (1+z)^4\nonumber\\ 
&&+ \Omega_{U0} (1+z)^{3(1+w_U)};
\ea
with $\Omega_{\Lambda 0}=1-\Omega_{m0}-\Omega_{U0}-\Omega_{r0}$. 
Here, $\Omega_{m0}, \Omega_{r0}, \Omega_{\Lambda0}, \Omega_{U0}$ 
are the relative abundance of matter, radiation, the 
cosmological constant, and unmatter respectively. Of course, 
for the $\rm \Lambda UDM$ model, $\Omega_{m0}=\Omega_{b0}$.

{\bf Supernova constraint}:
the luminosity distance to a supernova is given by 
\bq \label{fo:dl} d_L({z;\bf \theta}) = (1+z)\int_{0}^{z}\frac{cdz'}{H(z')}. 
\eq
and the distance moduli is 
\bq \label{eq:mut} 
\mu_{th}(z) = 5\log_{10}d_L(z) + 25,              
\eq
The $\chi^2$ for the SNIa data is 
\bq
\label{eq:chisq} \chi^2_{\rm SN}({\bf \theta}) = 
\sum_{i=1}^{N}\frac{(\mu_{obs}(z_i)-\mu_{th}(z_i))^2}{\sigma_i^2},
\eq
where $\mu_{obs}(z_i)$ and $\sigma_i$ are the observed value and the 
corresponding error for each supernova. We use a data set of 
182 high-quality SNIa \cite{Gong07} selected from 
the Gold06 \cite{Riess06}, SNLS \cite{Astier05} and ESSENCE
\cite{Wood07} samples.

{\bf CMB constraint:} the CMB shift parameter R \cite{Wang06} denotes the
positions of the acoustic peaks in the angular power spectrum of CMB, 
and takes the form as
\bq
{\rm R} = \sqrt{\Omega_{m0}}\int^{z_{\rm CMB}}_0\frac{dz'}{H(z')/H_0}
\eq
The WMAP3 data gives ${\rm R} = 1.70 \pm 0.03$ \cite{Wang07}, thus we have
\bq
\chi^2_{\rm R} = \Bigg(\frac{{\rm R}-1.70}{0.03}\Bigg)^2.
\eq

{\bf BAO constraint:} we use the quantity $\rm r_s/D_v$ which is constrained by 
the BAO signature in SDSS (at $z = 0.35$) and 2dFGRS (at $z = 0.2$) data
\cite{Percival07,WMAP08}: $\rm r_s/D_v(0.2) = 0.1980
\pm 0.0058$ and $\rm r_s/D_v(0.35) = 0.1094 \pm 0.0033$,
with a correlation coefficient of 0.39. Here 
$\rm r_s$ is the comoving sound horizon size at the epoch of decoupling, 
and $\rm D_v$ is the effective distance defined in \cite{Eisenstein05}.
we do not use the parameter $A$ which is extracted from the BAO measurements 
of the SDSS, as its definition applies to the $\Lambda$CDM model specifically, 
and may not be applicable in the presence of unmatter models \cite{Carneiro07}. 

For the combined analysis,
\bq
\chi^2 = \chi^2_{\rm SN} +  \chi^2_{\rm R}+ \chi^2_{\rm BAO} .
\eq

We employ the Markov Chain Monte Carlo (MCMC) technique to calculate the 
posterior probability distributions function of the parameters.
The Metropolis-Hastings algorithm with uniform priors is used to 
generate the sample, and the priors 
are taken as the following: $\Omega_{b0}\in(0, 0.1)$, 
$\Omega_{m0}\in(0, 1)$, $\Omega_{U0}\in(0, 1)$, 
$d_U\in(0, 10^5)$ and $h_0\in(0.4, 0.9)$.
The energy density of all components are assumed to be positive, 
$\Omega_{U0}\in(0, 1-\Omega_{b0}/\Omega_{m0}-\Omega_{r0})$ is set so that 
$\Omega_{\Lambda 0}=1-\Omega_{U0}-\Omega_{b0}/\Omega_{m0}-\Omega_{r0} \geq 0$. 
For each of the two models ($\Lambda$UDM and $\Lambda$CUDM)
we generate six chains, and about ten thousands points are sampled in 
each chain. After thinning the chains, we merge them into one chain 
which consists of about $10000$ points.

\section{Results}

First we consider the constraints derived purely from the SNIa
data. For constraining the unmatter model, this is the most
reliable one, as it is based only on the global expansion 
history, which can be calculated exactly for the given parameter set.

\begin{figure}[ht]
\begin{center}
\includegraphics[scale=0.3]{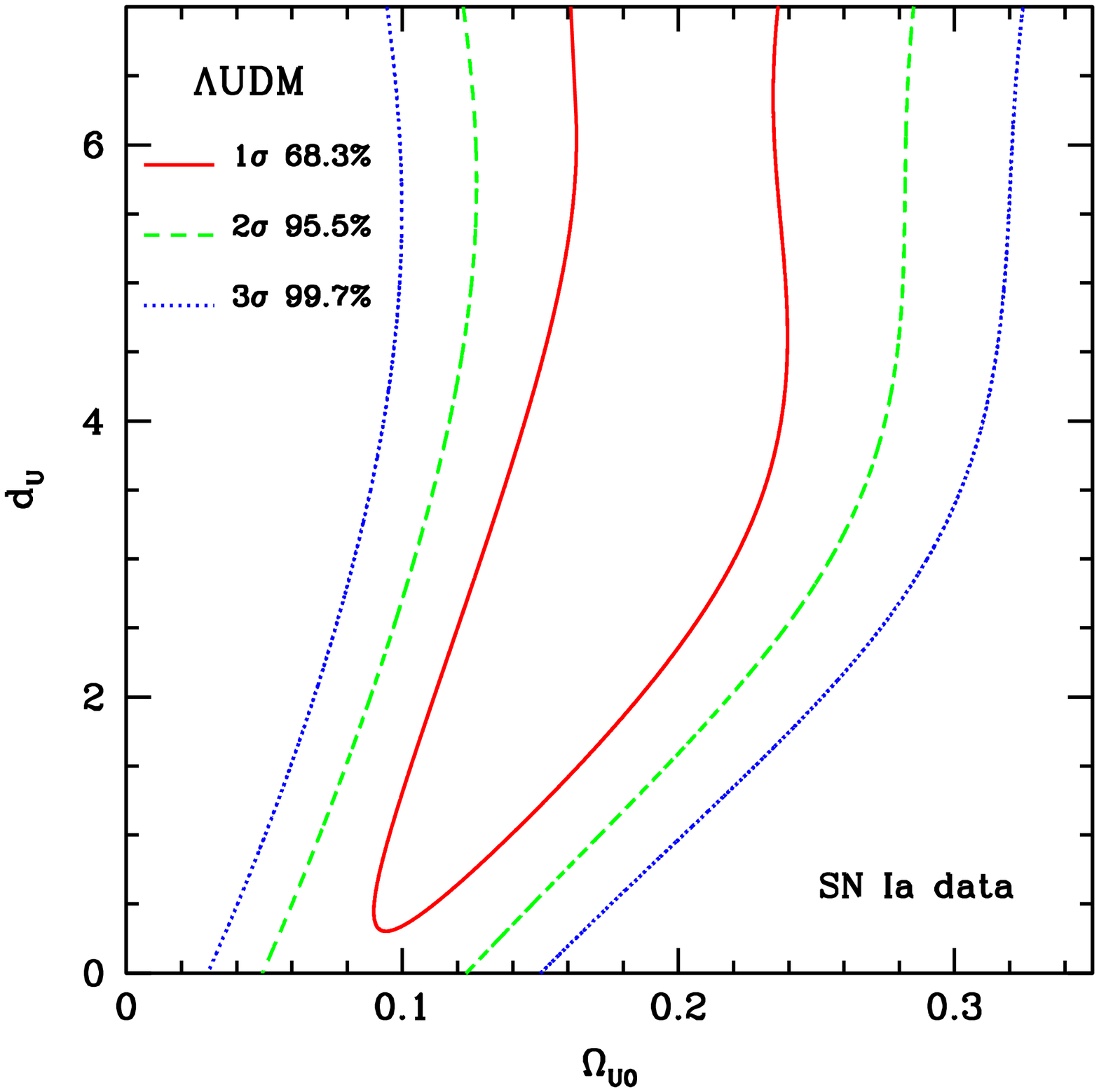}
\caption{\label{fig:SN_LUDM} The $1\sigma, 2\sigma, 3\sigma$ contours
of the $\Omega_U$ and $d_U$ in the $\Lambda$UDM model derived from
SNIa observations.}
\end{center}
\end{figure}

In Fig.~\ref{fig:SN_LUDM} we plot the constraint on $\Omega_U$ and 
$d_U$ in the $\Lambda$UDM model after marginalizing the other parameters.
We found that practically all values of $d_U$ are
allowed. At large values of $d_U$, the central value of $\Omega_U$ is 
between 0.15 and 0.25. This is what we would have expected, since for large value
of $d_U$ the behavior of the unparticle gas is very similar to that of the
cold dark matter, and for $\Lambda$CDM the best fit is centered in the same region.
At smaller values of $d_U$, the contours curved to smaller $\Omega_U$.

\begin{figure}[hb]
\begin{center}
\includegraphics[scale=0.3]{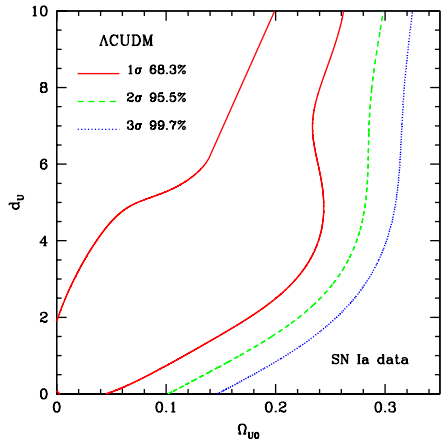}
\caption{\label{fig:SN_LCUDM} The $1\sigma, 2\sigma, 3\sigma$ contours
of the $\Omega_U$ and $d_U$ in the $\Lambda$CUDM model derived from
SN Ia observations.}
\end{center}
\end{figure}

In Fig.~\ref{fig:SN_LCUDM} we plot the constraint on $\Omega_U$
and $d_U$ in the $\Lambda$CUDM model. Again, practically all values of 
$d_U$ are allowed. For large $d_U$ where the unmatter asymptotes to cold 
dark matter, the best fit is located at $\Omega_U \sim 0.22$, as the UDM
become dominant and the CDM has a very small abundance. 
However, for high values of $d_U$, all values of $\Omega_U$ 
are allowed, in this part of the parameter space the UDM plays a minor
role and the CDM is dominant.

\begin{figure}[ht]
\begin{center}
\includegraphics[scale=0.3]{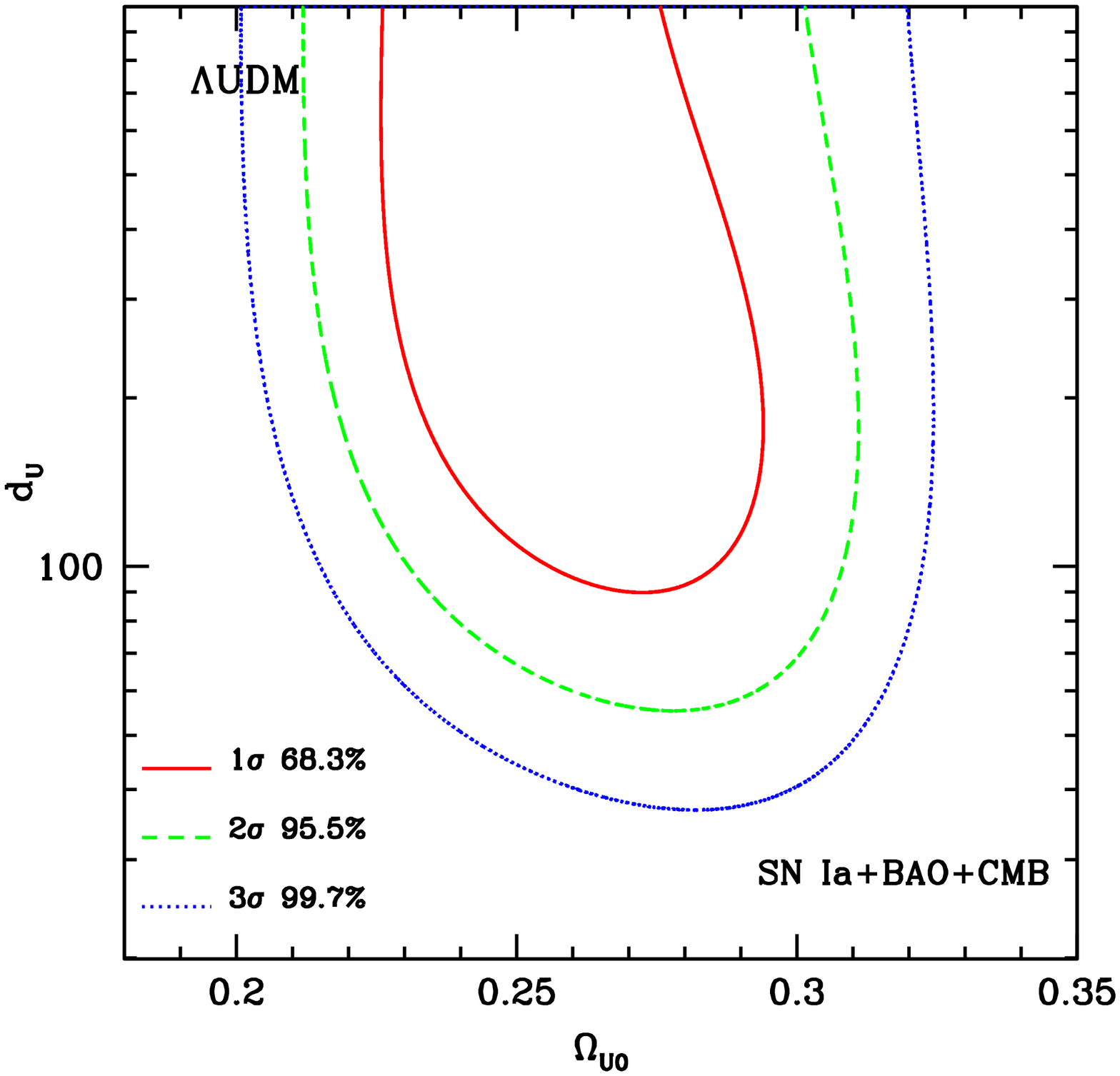}
\caption{\label{fig:SNR_LUDM} The $1\sigma, 2\sigma, 3\sigma$ contours
of the $\Omega_U$ and $d_U$ in the $\Lambda$UDM model derived from
SNIa, BAO and CMB shift parameter observations.}
\end{center}
\end{figure}

\begin{figure}
\begin{center}
\includegraphics[scale=0.3]{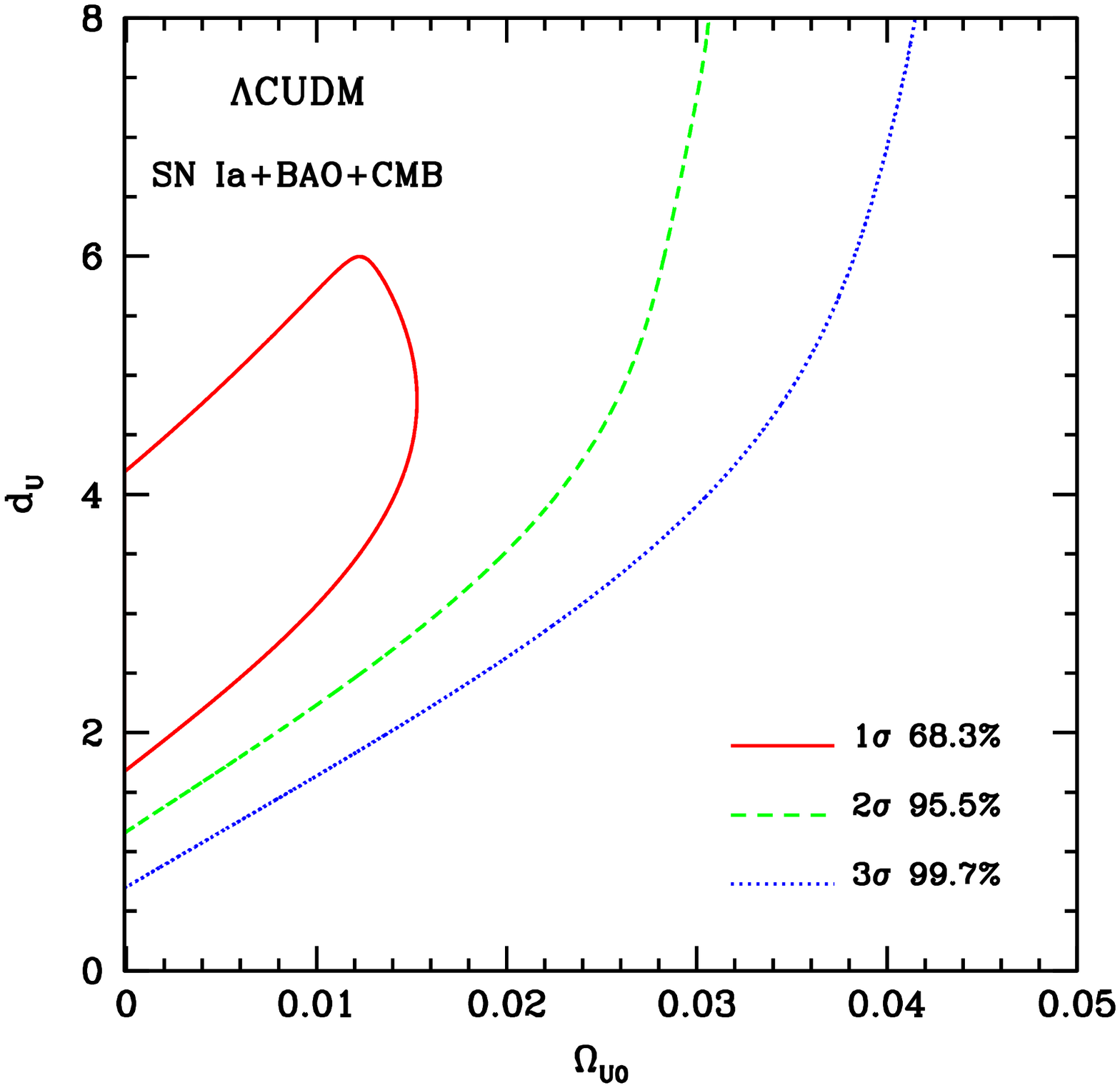}
\caption{\label{fig:SNR_LCUDM} The $1\sigma, 2\sigma, 3\sigma$ contours
of the $\Omega_U$ and $d_U$ in the $\Lambda$CUDM model derived from
SNIa, BAO and CMB shift parameter observations.}
\end{center}
\end{figure}

Our constraint is improved significantly by using the BAO and CMB shift 
parameter in addition to the SNIa data. The $\rm r_s/D_v$ proposed by
Percival et al. \cite{Percival07} is a ratio of two "standard ruler", 
which is suitable and reliable
to constrain our unmatter model and could efficiently break the degeneracy of
the parameters. The shift parameter does not include
all information in the CMB angular power spectrum, but it is relatively 
easy to compute for unconventional models. The large distance to the last scattering
surface of the CMB provides a long level arm for constraining the global
expansion rate. 
We plot the $\Omega_U-d_U$ contours for the $\Lambda$UDM model in 
Fig.~\ref{fig:SNR_LUDM}, and the same contours for the $\Lambda$CUDM
in Fig.~\ref{fig:SNR_LCUDM}. As can be seen from the figures, the
distribution is drastically different from the SNIa only constraints.

For the $\Lambda$UDM model, the lowest edge of the $2\sigma$ (95.5\%
C.L.) contour is about 60. Thus, small values of $d_U$, which is 
interesting from a physics perspective, is excluded at high
confidence levels. At large $d_U$ the unmatter is similar to cold dark
matter, so again it is not surprising that the center value of
$\Omega_U$ is 0.23-0.29.

For the $\Lambda$CUDM model the constraint
is also much more stringent. At $2\sigma$ level,
only 2 or 3 percent of the total cosmic density could be in unmatter.
For the scaling dimension $d_U$, the minimal allowed value is
1.2 at 95.5\%C.L.. Furthermore, for the small $d_U$ values, the constraint on the 
unmatter density is stronger. For $d_U<2$, the limit on unmatter
density is $\Omega_U<0.01$ at $2\sigma$ level. The constraint on abundance 
$\Omega_U$ loosened at greater values of $d_U$, where the unmatter
become indistinguishable from cold dark matter.

\section{Conclusion}

In this work, the scaling dimension $d_U$ and the abundance $\Omega_U$ of the
unparticle dark matter are constrained by using the SNIa luminosity 
distance moduli, BAO features in large scale structure as measured by 
SDSS and 2dFGRS, and the CMB shift parameter 
which depends on the cosmic expansion history.
We used the MCMC technique to simulate the posterior
probability distributions of the parameters.
Two models are considered, viz. $\Lambda$UDM (where unparticles are the only 
dark matter) and $\Lambda$CUDM (where unparticles co-exists with other cold dark 
matter).

Using only the supernova data, we find that the constraints on $d_U$
and $\Omega_U$ are pretty weak. However, with the addition of the BAO and
CMB shift parameter data, the degeneracy is broken, and strong constraint
could be put on $d_U$ and $\Omega_U$. For the $\Lambda$UDM model, $d_U>60$ at
$2\sigma$ level. For the $\Lambda$CUDM model, if $d_U<2$, then 
$\Omega_U<0.01$ at 95.5\% C.L.; and $\Omega_U$ is no greater than a few percent if
$d_U<10$. These limits severely constrained models of unparticle dark
matter and stable relic unparticle matter.

A major assumption adopted in this work is the equation of state 
$w_U$ for the unmatter: $w_U=1/(2d_U+1)$. This expression was
derived by S. L. Chen et al in Ref.~\cite{Chen07} for a thermal distribution 
of unparticle matter. This result is a little surprising, since an ideal gas of 
massless particles which also possesses scale invariance has an equation of state
$w_U=1/3$. However, if one accepts the usual density of states 
for the unparticles (Eq.~\ref{eq:dos}), which was used for other calculations 
about unparticles, e.g. production of unparticle in particle collisions,
then this equation of state can be derived using the standard method 
of statistical mechanics, as has been shown in 
Ref.~\cite{Chen07}. It is also possible that the unparticle 
matter has a non-thermal
distribution, then its equation of state would be different, and in that case
it is not constrained by our result. However, 
if the unparticle matter was produced during the hot Big Bang in the standard
decoupling scenario, its very nature must be considered a {\it thermal} 
distribution of unparticle matter.

\begin{acknowledgement}
We thank professors Miao Li, Xiaogang He and Yi Ling for discussions.  
Our MCMC chain computation was performed on the Supercomputing Center of 
the Chinese Academy of Sciences and the Shanghai Supercomputing
Center. This work is supported by
the National Science Foundation of China under the Distinguished Young
Scholar Grant 10525314, the Key Project Grant 10533010, by the
Chinese Academy of Sciences under the grant KJCX3-SYW-N2, and by the 
Ministry of Science and Technology national basic science
Program (Project 973) under grant No. 2007CB815401.  
X.C. also acknowledges the hospitality of the 
Moore Center of Theoretical Cosmology and Physics at Caltech and Kavli Institute
for Theoretical Physics in China, where part of this research was performed.   
\end{acknowledgement}

\newcommand\AL[3]{~Astron. Lett.{\bf ~#1}, #2~ (#3)}
\newcommand\AP[3]{~Astropart. Phys.{\bf ~#1}, #2~ (#3)}
\newcommand\AJ[3]{~Astron. J.{\bf ~#1}, #2~(#3)}
\newcommand\APJ[3]{~Astrophys. J.{\bf ~#1}, #2~ (#3)}
\newcommand\APJL[3]{~Astrophys. J. Lett. {\bf ~#1}, L#2~(#3)}
\newcommand\APJS[3]{~Astrophys. J. Suppl. Ser.{\bf ~#1}, #2~(#3)}
\newcommand\JCAP[3]{~JCAP. {\bf ~#1}, #2~ (#3)}
\newcommand\JHEP[3]{~JHEP. {\bf ~#1}, #2~ (#3)}
\newcommand\LRR[3]{~Living Rev. Relativity. {\bf ~#1}, #2~ (#3)}
\newcommand\MNRAS[3]{~Mon. Not. R. Astron. Soc.{\bf ~#1}, #2~(#3)}
\newcommand\MNRASL[3]{~Mon. Not. R. Astron. Soc.{\bf ~#1}, L#2~(#3)}
\newcommand\NPB[3]{~Nucl. Phys. B{\bf ~#1}, #2~(#3)}
\newcommand\PLB[3]{~Phys. Lett. B{\bf ~#1}, #2~(#3)}
\newcommand\PRL[3]{~Phys. Rev. Lett.{\bf ~#1}, #2~(#3)}
\newcommand\PR[3]{~Phys. Rep.{\bf ~#1}, #2~(#3)}
\newcommand\PRD[3]{~Phys. Rev. D{\bf ~#1}, #2~(#3)}
\newcommand\SJNP[3]{~Sov. J. Nucl. Phys.{\bf ~#1}, #2~(#3)}
\newcommand\ZPC[3]{~Z. Phys. C{\bf ~#1}, #2~(#3)}

\end{document}